\documentclass[11pt]{article}
\def\half{\frac{1}{2}}

\newfont{\bbbold}{msbm10 scaled \magstep1}

\def\bbR{\mbox{\bbbold R}}

\def\bbZ{\mbox{\bbbold Z}}
\def\cA{{\cal A}}

\def\cF{{\cal F}}

\def\cL{{\cal L}}
\def\cM{{\cal M}}

\def\cU{{\cal U}}

\newfont{\goth}{eufm10 scaled \magstep1}

\def\gi{\mbox{\goth i}}

\def\gn{\mbox{\goth n}}

\def\gp{\mbox{\goth p}}

\def\gs{\mbox{\goth s}}

\def\a{\alpha}
\def\b{\beta}
\def\c{\gamma}\def\C{\Gamma}
\def\d{\delta}
\def\e{\epsilon}\def\ve{\varepsilon}
\def\f{\phi}
\def\h{\eta}
\def\i{\iota}
\def\k{\kappa}
\def\l{\lambda}\def\L{\Lambda}

\def\r{\rho}
\def\s{\sigma}
\def\t{\tau}\def\tt{\tilde{\tau}}
\def\th{\theta}

\def\be{\begin{equation}}\def\ee{\end{equation}}
\def\bea{\begin{eqnarray}}\def\eea{\end{eqnarray}}
\def\barr{\begin{array}}\def\earr{\end{array}}

\def\O{\Omega}
\def\del{\partial}
\def\ua{\underline{\alpha}}

\def\una{\underline a}\def\unA{\underline A}
\def\unb{\underline b}\def\unB{\underline B}
\def\unC{\underline C}

\def\unM{\underline M}

\def\xz{\times}

\def\nab{\nabla}

\def\tq{\widetilde{q}}\def\tv{\widetilde{v}}

\def\tV{\widetilde{V}}

\def\tL{\widetilde{\L}}
 
\let\la=\label

\def\nn{\nonumber}
\def\bd{\begin{document}}
\def\ed{\end{document}}
\def\ba{\begin{array}}
\def\ea{\end{array}}
\def\bea{\begin{eqnarray}}
\def\eea{\end{eqnarray}}
\def\ft#1#2{{\textstyle{{\scriptstyle #1}\over {\scriptstyle #2}}}}
\def\fft#1#2{{#1 \over #2}}
\newcommand{\eq}[1]{(\ref{#1})}
\newcommand{\w}[1]{\\[0.#1cm]}
\def\eqs#1#2{(\ref{#1}-\ref{#2})}
\def\det{{\rm det\,}}
\def\tr{{\rm tr}}
\newcommand{\hoch}[1]{$\, ^{#1}$}
\newcommand{\tamphys}{\it\small Center for Theoretical Physics,
Texas A\&M University, College Station, TX 77843, USA}
\newcommand{\kings}
{\it\small Department of Mathematics, King's College, Strand, London WC2R 2LS, UK}
\newcommand{\gron}
{\it\small Centre for Theoretical Physics, University of Groningen, Nijenborgh 4, 9747 AG Groningen,
The Netherlands}
\newcommand{\nik}
{\it\small Research \& Development, Nik Software GmbH, Hammerbrookstr. 93, Hamburg, Germany}
\newcommand{\galileo}
{\it\small Dipartimento di Fisica "Galileo Galilei", Universit\`a degli Studi di Padova \& INFN,
Sezione di Padua, via F. Marzolo 8, 35131 Padova, Italia}
\newcommand{\stock}
{\it\small Department of Theoretical Physics, Stockholm, Sweden}
\makeatletter
\renewcommand\theequation{\thesection.\arabic{equation}}
\@addtoreset{equation}{section} \makeatother

\newcommand{\auth}
{\large E. Bergshoeff\hoch{1}, P.S. Howe\hoch{2}, S. Kerstan\hoch{3} and L. Wulff\hoch{4}}

\begin{document}

\hfill{KCL-TH-07-12}

\hfill{UG-07-05}

\vspace{20pt}

\begin{center}

{\Large{\bf Kappa-symmetric $SL(2,\bbR)$ covariant D-brane actions}}

\vspace{30pt}

\auth

\vspace{15pt}

\begin{itemize}
\item [$^1$] \gron \item [$^2$] \kings \item[$^3$] \nik
\item  [$^4$] \galileo
\end{itemize}

\vspace{60pt}

{\bf Abstract}

\end{center}

A superspace formulation of IIB supergravity which includes the field strengths of the duals of the usual physical one, three and five-form field strengths as well as the eleven-form field strength is given. The superembedding formalism is used to construct kappa-symmetric $SL(2,\bbR)$ covariant D-brane actions in an arbitrary supergravity background.

\vskip 6cm

\rule{14cm}{.05mm}

{\sl E.A.Bergshoeff@rug.nl, paul.howe@kcl.ac.uk, s.f.kerstan.99@cantab.net, linus@pd.infn.it}

\pagebreak \tableofcontents \setcounter{page}{1}


\section{Introduction}


IIB supergravity was written down in components in \cite{Schwarz:1983qr,Schwarz:1983wa} and in superspace in \cite{Howe:1983sr}. As is well-known, the bosonic fields are the graviton, two spin 0 fields, two two-form potentials and a four-form potential with a self-dual five-form field strength. The duals of these are important for couplings to branes; a superspace formulation including these extra fields was given in \cite{Dall'Agata:1998va}, while the D-brane actions were written down in supergravity backgrounds in \cite{Cederwall:1996ri,Bergshoeff:1996tu}. Recently \cite{Bergshoeff:2005ac}, it has been shown that there are also a number of ten-form potentials, transforming under the doublet and quartet representations of $SL(2,\bbR)$, whose presence is in accordance with extended symmetry considerations \cite{Kleinschmidt:2003mf}. In this paper we first extend the superspace formalism to include the ten-form potentials and then go on to use it to construct kappa-symmetric $SL(2,\bbR)$ covariant actions for D-branes in arbitrary IIB supergravity backgrounds.

The superspace formalism has some advantages over the component formalism for discussing the ten-form potentials. This is because, in superspace, the associated eleven-form field strengths do not vanish so that it is possible to give a gauge-invariant discussion with manifest supersymmetry. We shall write down the components of all the superspace form field strengths in the next section, including the doublet of eleven-forms which is not needed for the D-branes under discussion here. We use a formalism with local $SO(2)$ and global $SL(2,\bbR)$ symmetries.

Duality-symmetric actions have been given for strings \cite{Townsend:1997kr}, superstrings \cite{Cederwall:1997ts} and three-branes \cite{Berman:1997iz,Cederwall:1997ab,Berman:1998va,Nurmagambetov:1998gp}, the latter with the aid of the results of \cite{Pasti:1995tn,Pasti:1996vs}, while an attempt to carry out a similar construction for five-branes was made in \cite{Westerberg:1999fe}. More recently, $SL(2,\bbR)$ invariant actions for the bosonic sector of all D-branes have been derived \cite{Bergshoeff:2006gs}.
The remainder of the current paper is devoted to the construction of kappa-symmetric $SL(2,\bbR)$ covariant actions for D-branes using the superembedding formalism \cite{Sorokin:1989zi,Delduc:1992fk,Bandos:1995zw}. This was first applied to D-branes in \cite{Howe:1996mx}. It will be shown, given the standard superembedding constraints, that supersymmetry completely determines the dynamics of D-branes and the associated actions.


\section{IIB supergeometry}

We begin by discussing the superspace geometry of IIB supergravity including the eleven-form field
strengths implied by the results of \cite{Bergshoeff:2005ac}. We use a real basis for the spinors, $E^{\a i}$, where
$\a=1,\ldots 16$ is a chiral ten-dimensional spinor index, while $i=1,2$ is a $Spin(2)$ spinor
index. The connection, $\O$, and curvature, $R$, correspondingly take their values in
$\gs\gp\gi\gn(1,9)\oplus\gs\gp\gi\gn(2)$. For example,

\be
 \O_{\a i}{}^{\b j}=\d_i{}^j \O_\a{}^\b + \d_\a{}^\b \ve_i{}^j Q\ ,
 \la{2.1}
\ee

where $Q$ is the $\gs\gp\gi\gn(2)$ connection and

\be
 \O_\a{}^\b=\frac{1}{4}(\c^{ab})_\a{}^\b \O_{ab}\ ,
 \la{2.1.1}
\ee

with $\O_{ab}$ being the Lorentz or $\gs\gp\gi\gn(1,9)$ connection. Small latin indices from the beginning of the alphabet denote Lorentz vector indices as usual.

We shall also use $SO(2)$ vector indices which will be
denoted by $r,s$, etc, while $SL(2,\bbR)$ doublet indices will be denoted $R,S$, etc. The scalar
fields $\cU$ carry indices $\cU_r{}^R$, i.e. (local) $SO(2)$ acts to the left and (rigid)
$SL(2,\bbR)$ to the right. The dimension zero torsion is

\be
 T_{\a i\b j}{}^c=-i\d_{ij}(\c^c)_{\a\b}\ .
 \la{2.2}
\ee

The non-zero dimension one-half torsion is

\be
 T_{\a i\b j}{}^{\c k}=-i\left((\c^a)_{\a\b} (\c_a)^{\c\d}-2
 \d_{(\a}{}^\c\d_{\b)}{}^\d\right)\L_{\d ij}{}^k\ ,
 \la{2.2.1}
\ee

where $\L_{ijk}$ is totally symmetric and traceless\footnote{We suppress the spinor index on $\L$ in the text.}.

The forms consist of a triplet of one-forms, a doublet of three-forms, a singlet five-form, a
doublet of seven-forms, a triplet of nine-forms and a quadruplet of eleven-forms. There is also a
doublet of eleven-forms which does not feature in the brane actions we shall consider here. These
can be written as $SL(2,\bbR)$ representations or as $SO(2)$ representations, the two being related
by the scalar matrix $\cU$. For example, the three-form is a doublet $F_R$, and $F_r=\cU_r{}^R
F_R$. We can also use $\cU$ to define a metric by

\be
 \cM_{RS}:=(\cU^{-1})_R{}^r (\cU^{-1})_S{}^s \d_{rs}\ .
 \la{2.3}
\ee

The derivative of $\cU$ is given by

 \be
 (d\cU \cU^{-1})_r{}^s=(F^{(1)})_r{}^s +2 \ve_r{}^s Q
 \la{2.4}
 \ee

where $Q$ is the $U(1)$ connection and $F^{(1)}_{rs}$ is symmetric and traceless, $\d^{rs}
F^{(1)}_{rs}=0$.

In this notation the Bianchis are (with the form ranks as
superscripts)

 \begin{eqnarray}
   d F^{(1)}_{RS}&=& 0\nn\w1
   d F^{(3)}_R &=& 0\nn \w1
   d F^{(5)} &=&- \ve^{RS} F^{(3)}_R F^{(3)}_S\nn\w1
   d F^{(7)}_R &=& F^{(3)}_R F^{(5)}\nn\w1
   d F^{(9)}_{RS} &=&  F^{(3)}_{(R} F^{(7)}_{S)}\nn\w1
   d F^{(11)}_{RST}&=&F^{(3)}_{(R}F^{(9)}_{ST)} \ .
   \la{2.5}
 \end{eqnarray}

Note that the tracelessness condition for $F^{(1)}$ becomes $\cM^{RS} F^{(1)}_{RS}=0$ in the
$SL(2,\bbR)$ basis; in fact, one can show that $d\cM_{RS}=-2F^{(1)}_{RS}$. It is straightforward to
rewrite the Bianchis in the $SO(2)$ basis; for each index one gets a factor of $F^{(1)}$. For
example,

\be
 DF^{(3)}_{r}=-(F^{(1)})_r{}^s F^{(3)}_s\ .
 \la{2.6}
\ee

The dimension zero components of the forms, in $SO(2)$ notation, are

\bea
  F^{(3)}_{r\a i\b j c}&=&-i(\t_r)_{ij}(\c_c)_{\a\b} \nn\w1
  F^{(5)}_{\a i\b j cde}&=& i\ve_{ij}(\c_{cde})_{\a\b}\nn\w1
  F^{(7)}_{r\a i\b j c_1\ldots c_5}&=& i(\tt_r)_{ij}(\c_{c_1\ldots c_5})_{\a\b}\nn\w1
  F^{(9)}_{rs\a i\b j c_1\ldots c_7}&=& \frac{i}{2} \ve_{ij}\d_{rs}(\c_{c_1\ldots c_7})_{\a\b}\nn\w1
  F^{(11)}_{rst \a i\b j c_1\ldots c_9} &=& \frac{i}{2} \d_{(rs}(\tt_{t)})_{ij}
  (\c_{c_1\ldots c_9})_{\a\b}\ ,
  \la{2.7}
\eea

where $\t_r=\frac{1}{\sqrt{2}}(\s_3,\s_1)$ are the $SO(2)$ gamma-matrices and $\tt_r=\ve_{rs}\t^s$.
The other components of the forms are not needed for the D-branes but we shall give them here for
completeness.

At dimension one-half we have

\bea
 F^{(1)}_{rs \a i}&=&-2i(\t_r \L_s)_{i\a}\nn \w1
 F^{(3)}_{r\a i b_1 b_2} &=&-i  \left(\c_{b_1 b_2} \L_{r i}\right)_\a \nn \w1
 F^{(5)}_{\a i b_1\ldots b_4} &=& 0 \nn \w1
 F^{(7)}_{r \a i b_1\ldots b_6}&=&-i \left(\c_{b_1\ldots b_6} \tL_{r i}\right)_\a \nn \w1
 F^{(9)}_{rs \a i b_1\ldots b_8} &=&-2i\left(\c_{b_1\ldots b_8}\t_{(r} \tL_{s)}\right)_{i \a} \nn \w1
 F^{(11)}_{rst \a i b_1\ldots b_{10}} &=&-\frac{3i}{2}  \left(\c_{b_1\ldots b_{10}} \d_{(rs}\tL_{t) i}\right)_\a\ .
 \la{2.8}
\eea

The field $\L_{ri}$ is the dimension one half spinor field; it can be written

\be
 \L_{r i}= (\t_r)^{jk} \L_{ijk}
 \la{2.10}
\ee

where $\L_{ijk}$ is the field appearing in the dimension one-half torsion.

We have also defined

\be
 \tL_{r i}:=\ve_{rs} \L^s{}_{i}=\ve_{ij} \L_{r}{}^{j}\ .
 \la{2.12}
\ee

The dimension one components of the one-, three- and five-form field strengths are the superfields
whose leading components are the (covariantised) spacetime field strengths (with $\d^{rs}
F^{(1)}_{rs a}=0$). The seven-form field strengths are essentially the duals of the three-forms,

\be
 F^{(7)}_{r a_1 \ldots a_7}= \frac{1}{3!}\ve_{rs} \ve_{a_1\ldots a_7 b_1 b_2 b_3} F^{(3)s b_1 b_2 b_3}\ ,
 \la{2.13}
\ee

while the nine-form dimension one components are

\be
 F^{(9)}_{rs a_1\ldots a_9}=-\ve_r{}^t \ve_{a_1\ldots a_9 b} F_{st}^{(1) b} -3i\ve^{ij}
 \L_{r i}\c_{a_1\ldots a_9}\L_{s j} \ .
 \la{2.14}
\ee

The eleven-forms are of course identically zero at dimension one, while the five-form is self-dual up to non-linear terms,

\be
 F^{(5)}_{a_1\ldots a_5}=\frac{1}{5!}\ve_{a_1\ldots a_5 b_1\ldots b_5} F^{(5)b_1\ldots b_5}+ i\d^{rs}\ve^{ij}\L_{r i}  \c_{a_1\ldots a_5}\L_{s j}\ .
 \la{2.14.1}
\ee

The non-vanishing components of the eleven-form doublet, $F^{(11)}_r$, are

\bea
 F^{(11)}_{r \a i\b j c_1\ldots c_9}&=& i(\t_r)_{ij}(\c_{c_1\ldots c_9})_{\a\b}\nn\w1
 F^{(11)}_{r \a i c_1\ldots c_{10}}&=&i\frac{27}{23} (\c_{c_1\ldots c_{10}} \L_{ri})_\a\ .
 \la{2.15}
\eea

This form satisfies the Bianchi identity

\be
 dF^{(11)}_R=\frac{4}{23}\left( \ve^{ST} F^{(3)}_S F^{(9)}_{TR}-\frac{3}{4} F^{(5)}F^{(7)}_R
 \right)\
 \la{2.16}
\ee

in the $SL(2,\bbR)$ basis. This result implies that the doublet of ten-form potentials transforms under the gauge transformations of some of the other $p$-form potentials, in agreement with the extended symmetry considerations of \cite{Kleinschmidt:2003mf}. In \cite{Bergshoeff:2005ac} these additional transformations are not present, but this is not a contradiction since any ten-form gauge transformation can be written as the exterior derivative of a nine-form gauge transformation in ten-dimensional spacetime.\footnote{We thank Mees de Roo and Diederik Roost for a discussion of this point.}


\section{Superembeddings}



\subsection{Charges}


In order to discuss branes and superembeddings in an $SL(2,\bbR)$ covariant way we follow reference \cite{Bergshoeff:2006gs} and
introduce a pair of constant charge vectors $(q^R,\tq^R)$. They can be normalised so that

\be
 q^R \ve_{RS} \tq^S =2\ .
 \la{3.1}
\ee

We define $q^r:=q^R (\cU^{-1})_R{}^r$ and $\tq^r:=\tq^R (\cU^{-1})_R{}^r$, although note that $\tq^r$ is not the dual of $q^r$. Let $(V^r,\tV^r)$, $\tV^r:=\ve^{rs}
V_s$, be an orthonormal pair of vectors\footnote{Note that the $V^r$, as well as the $v^i$ introduced in \eq{3.4}, are not constant.}; they can be chosen such that

\bea
 q^r&=& a V^r \nn\w1
 \tq^r&=& -\frac{2}{a}\tV^r + b V^r\ ,
 \la{3.2}
\eea

where $a$ and $b$ are functions of the scalars given by

\bea
 \cM_{RS} q^R q^S&=& a^2 \nn\w1
 \cM_{RS} \tq^R \tq^S&=&b^2 + \frac{4}{a^2}\ .
 \la{3.3}
\eea

The $b$-term in the expression for $\tq^r$ is necessary because $q^R$ and $\tq^R$ are constant. We
also introduce an orthonormal pair of $Spin(2)$ vectors $(v^i,\tv^i),\ \tv^i:=\ve^{ij} v_j$, such
that

\bea
 V_r &=& \sqrt{2} (\t_r)_{ij} v^i v^j \nn\w1
 \tV_r &=& \sqrt{2} (\t_r)_{ij} v^i \tv^j\ .
 \la{3.4}
\eea

In the purely bosonic sector of the theory it is not necessary to introduce the unit vectors $v^i$ and $V^r$\cite{Bergshoeff:2006gs}, but it is useful in the superembedding context as we shall see.


\subsection{The superembedding matrix}


In the superembedding formalism the worldvolume of the brane is a superspace $M$ whose odd
dimension is half that of the target space $\unM$. Super, even, odd indices are respectively
denoted by capital, latin, greek letters. Letters from the beginning of the alphabet refer to
preferred bases while indices form the middle refer to coordinate bases. Indices for the target
space are underlined while normal indices are primed. The preferred coframes are denoted by
$E^A=(E^a,E^\a)$, where $E^A=dz^M E_M{}^A$, and similarly the frames are $E_A=E_A{}^M {\del_M}$
where $E_A{}^M$ is the inverse of the supervielbein $E_M{}^A$. The embedding matrix is the
derivative of the embedding map with respect to the preferred basis,

\be
 E_A{}^{\unA}:=E_A{}^M \del_M z^{\unM} E_{\unM}{}^{\unA}\ .
 \la{3.5}
\ee

The embedding constraint is

\be
 E_{\a}{}^{\una}=0\ .
 \la{3.6}
\ee

For most branes this implies the equations of motion. For D-branes there is a worldvolume gauge
field $\cA$ whose modified field strength $\cF$ obeys the constraint

\be
 \cF_{\a B}=0\ .
 \la{3.7}
\ee

These two constraints are always imposed and imply the equations of motion. In the IIB case we
write $E^{\ua}=E^{\a i}$; the remainder of the superembedding matrix can then be parametrised in
the following way:

\bea
 E_a{}^{\unb}&=& u_a{}^{\unb}\nn\w1
 E_{\a}{}^{\b j}&=& u_{\a}{}^{\b} v^j + h_{\a}{}^{\c} u_{\c}{}^{\b} \tv^j\nn\w1
 E_a{}^{\b j}&=&\l_a{}^{\c} u_{\c}{}^{\b} \tv^j\ .
 \la{3.8}
\eea

where $u_\a{}^{\b}$ is an element of $Spin(1,9)$, $(u_a{}^{\unb},u_{a'}{}^{\unb})$ together give
the corresponding element of the Lorentz group, and where $\l_a{}^{\b}$ can be thought of as the
bosonic derivative of the transverse fermions in the brane multiplet. The field $h_\a{}^\b$ is
related to $\cF$ in a non-linear fashion to be discussed below.


\subsection{Torsion equation}


In order to work out the consequences of the superembedding constraints we shall need the torsion
equation:

\be
 2\nab_{[A} E_{B]}{}^{\unC} + T_{AB}{}^C E_C{}^{\unC}= (-1)^{A(B+\unB)} E_B{}^{\unB} E_A{}^{\unA}
 T_{\unA\unB}{}^{\unC}\ ,
 \la{3.9}
\ee

and the Bianchi identity for $\cF$, $d\cF=-H$, where $H$ is a target space three-form to be defined
shortly. In components, this Bianchi identity is

\be
 3\left( \nab_{[A} \cF_{BC]} + T_{[AB}{}^D \cF_{|D|C]}\right)=- H_{ABC}\ ,
 \la{3.10}
\ee

where $H$ here is pulled back from the target space to the worldvolume using the embedding matrix.

We define $H$ to be

\be
 H:=q^R F^{(3)}_R=q^r F^{(3)}_r\ .
 \la{3.11}
\ee

The dimension zero component of the torsion equation, projected along the brane, implies

\be
 T_{\a\b}{}^c=-i(\c^c + h\c^c h^T)_{\a\b}\ ,
 \la{3.14}
\ee

while the normal projection gives

\be
 (\c^{c'} + h\c^{c'} h^T)_{\a\b}=0\ .
 \la{3.15}
\ee

The dimension zero component of the $\cF$ Bianchi gives

\be
 T_{\a \b}{}^d \cF_{dc}=ia'(\c^c - h\c^c h^T)_{\a\b};\qquad a':=   \frac{a}{\sqrt{2}}
 \la{3.16}
\ee

which, together with \eq{3.14}, gives the relation between $h$ and $\cF$:

\be
 h\c^a h^T=\c^b L_b{}^a\ ,
 \la{3.17}
\ee

where

\be
 L_a{}^b:=\left((1+\cF')(1-\cF')^{-1}\right)_a{}^b\ ,\qquad \cF'_{ab}:=\frac{1}{a'}\cF_{ab}\ ,
 \la{3.18}
\ee

is an element of the Lorentz group $SO(1,p)$.\footnote{This relation between $h$ and $\cF$ was first observed for the D9-brane in \cite{Akulov:1998bq}; it is discussed for a general IIB D-brane in \cite{Bandos:2006wb}.}

These equations are solved by

\be
 h=h_0 \c_{(p+1)}\ ,
 \la{3.19}
\ee

where

\be
 \c_{(p+1)}:=\frac{1}{(p+1)!}\ve_{a_1\ldots a_{p+1}} \c^{a_1\ldots a_{p+1}}\ ,
 \la{3.20}
\ee

and where $h_0$ is an element of $Spin(1,p)$ corresponding to $L$. Explicitly \cite{Callan:1988wz},

\be
 h_0= \frac{1}{L_0}\sum\, \frac{1}{2^m m!} \c^{a_1b_1\ldots a_m b_m} \cF'_{a_1 b_1}\ldots
 \cF'_{a_m b_m}\ ,
 \la{3.21}
\ee

with

\be
 L_0=\sqrt{-\det\, (\h_{ab} + \cF'_{ab})}\ .
 \la{3.22}
\ee


\section{D-brane actions}



\subsection{General construction}


The GS action for a brane can be constructed from the superembedding formalism using the following
recipe \cite{Howe:1998ts} (see \cite{Bandos:1995dw}
for a related approach which was applied to D-branes in \cite{Bandos:1997rq}). For each $p$-brane there is a closed $(p+2)$-form $W$ which can be written as $dL_{WZ}$,
where $L_{WZ}$ is the Wess-Zumino term regarded as a $(p+1)$-form, and also as $dK$ where $K$ is a
tensorial $d$-form on the brane ($d=p+1$). Therefore $L_d=K-L_{WZ}$ is a closed $d$-form on the brane
which can be used to construct the action using the superform (ectoplasm) method \cite{Gates:1997ag}. The action is

\be
 S=\int_{M_o} \cL
 \la{3.23}
\ee

where $M_o$ is the body of $M$, i.e. the usual bosonic worldvolume, and

\be
 \cL:=\frac{1}{(p+1)!} d x^{m_d}\ldots dx^{m_1} L_{m_1\ldots m_d}(x,0)\ .
\ee

The construction guarantees that the action is invariant under local supersymmetry tranformations
on the brane, i.e. kappa-symmetry, and also under reparametrisations  of $M_0$. This can be seen as
follows: under an infinitesimal diffeomorphism of $M$ generated by a vector field $X$ we have

\be
 \d L_d = d\i_X L_d + i_X d L_d\ .
 \la{3.24}
\ee

Evaluating this equation at $\th=0$ and using the fact that $d L_d=0$ we see that $\cL$ will
transform as a total derivative under such a transformation. We can identify the even and odd
leading components of $X$ as the parameters of worldvolume diffeomorphisms and kappa-symmetry
respectively. For kappa-symmetry,

\be
 X=\k^\a E_\a=\k^\a E_\a{}^{\ua} E_{\ua}:=\k^{\ua} E_{\ua}\ .
 \la{3.24.1}
\ee

The above definition of $\k^{\ua}$ ensures that it satisfies $\k=\k P$ where $P$ is the projector
from the odd tangent space of the target superspace onto the odd tangent space of the brane. As $P$
is a projector it can be written as $P=\half (1 + \C)$ where $\C^2=1$, so that $\k=\half\k (1+\C)$.

To show that $W$ is exact it is convenient to introduce the notion of an $(r,s)$ form, one which
has $r$ even and $s$ odd indices with respect to a preferred basis \cite{Bonora:1990mt}. The exterior derivative can be
split into four parts $d_0\,,d_1\,,t_0\,,t_1$ with respective bidigrees
$(1,0),(0,1),(-1,2),(2,-1)$. $d_0$ and $d_1$ are even and odd derivatives, although they include
torsion components as well, while $t_0$ and $t_1$ are algebraic operations involving the dimension
zero and three-halves components of the torsion. From $d^2=0$ we find

\be
 t_0^2= d_1 t_0 + t_0 d_1=0\ ,
 \la{3.25}
\ee

together with some other equations which we shall not need. Since $t_0^2=0$ there are cohomology
groups $H_t^{r,s}$ whose elements are $(r,s)$ forms which are $t_0$-closed but not exact \cite{Bonora:1990mt}.

The lowest non-vanishing component of $W$ is $W_{p,2}$; as $dW=0$, we have $t_0 W_{p,2}=0$. It is
not difficult to show (see Appendix B) that $H_t^{p,2}=0$ from which we deduce that $W_{p,2}=t_0 K_{p+1,0}$ for some
$K_{p+1,0}$. The only other non-vanishing component of $W$ is $W_{p+1,1}$; it satisfies

\be
 d_1 W_{p,2} + t_0 W_{p+1,1}=0\ .
 \la{3.26}
\ee

Since $W_{p,2}=t_0 K_{p+1,0}$ and $d_1 t_0 + t_0 d_1=0$ we find

\be
 t_0 (W_{p+1,1}-d_1 K_{p+1,0})=0\ .
 \la{3.27}
\ee

It is straightforward to see that there are no non-trivial solutions to this equation from which we
conclude that

\be
 W_{p+1,1}=d_1 K_{p+1,0}\ .
 \la{3.28}
\ee

We have therefore shown that, if the lowest non-vanishing component of $W$ is $W_{p,2}$ then
$W=dK$, where $K=K_{p+1,0}$. If we can construct a suitable closed $(p+2)$ form $W$ we will
automatically have shown that there is a corresponding GS action. To complete the picture we shall
therefore only have to evaluate the dimension zero component of $W=dK$ to show that $K$ is indeed
the Dirac-Born-Infeld form, $L_{DBI}$.


\subsection{RR forms}


The Wess-Zumino form for a D$p$-brane is given by the $(p+2)$-form component of

\be
 W=e^{-\cF}\sum_n\, G^{(2n+1)}\ ,
 \la{4.1}
\ee

where the ``RR'' forms $G$ are pull-backs of forms on the target space. They satisfy the Bianchi
identities

\be
 dG^{(2n+1)}=H G^{(2n-1)}\
 \la{4.2}
\ee

which ensure that $dW=0$. They can be written in terms of potentials as

\be
 G^{(2n+1)}= d C^{(2n)} + H C^{(2n-2)}\ ,
 \la{5.11}
\ee

The forms are as follows:

\bea
 G^{(1)}&=& d\left(\frac{b}{a}\right) \nn\w1
 G^{(3)}&=&-\tq^r F^{(3)}_r + \frac{b}{a}H\nn\w1
 G^{(5)}&=& F^{(5)} \nn\w1
 G^{(7)}&=& q^r F^{(7)}_r\nn\w1
 G^{(9)}&=& q^r q^s F^{(9)}_{rs}\nn\w1
 G^{(11)}&=& q^r q^s q^t F^{(11)}_{rst}\ .
 \la{4.3}
\eea

We can read off the dimension zero components straightforwardly from \eq{2.7}. They are

\bea
  G^{(3)}_{\a i\b j c}&=&\frac{2i}{a}V^r(\tt_r)_{ij}(\c_c)_{\a\b} \nn\w1
  G^{(5)}_{\a i\b j cde}&=& i\ve_{ij}(\c_{cde})_{\a\b}\nn\w1
  G^{(7)}_{\a i\b j c_1\ldots c_5}&=& iaV^r(\tt_r)_{ij}(\c_{c_1\ldots c_5})_{\a\b}\nn\w1
  G^{(9)}_{\a i\b j c_1\ldots c_7}&=& \frac{ia^2}{2} \ve_{ij}(\c_{c_1\ldots c_7})_{\a\b}\nn\w1
  G^{(11)}_{\a i\b j c_1\ldots c_9} &=& \frac{ia^3}{2} V^r(\tt_{r})_{ij}
  (\c_{c_1\ldots c_9})_{\a\b}\ .
  \la{4.4}
\eea

If one sets $q^r=e^{\frac{\f}{2}}(\sqrt{2},0)$, where $\f$ is the dilaton, one recovers the
standard form for the dimension zero RR fields in the Einstein frame,

\be
 (G^{(2n+1)})_{2n-1,2}=ie^{(n-2)\frac{\f}{2}}E^{\b 2} E^{\a 1} (\c^{(2n-1)})_{\a\b}\ .
 \la{4.5}
\ee


\subsection{Kappa-symmetry}


According to the general argument given previously, the closed $(p+2)$-form $W$ gives rise to the
GS action, and the latter is automatically kappa-symmetric. In this section we verify explicitly
that $K=L_{DBI}$, where

\be
 L_{DBI}=f L_0\, \ve_{(p+1)}\ .
 \la{5.1}
\ee

The function $f$ is an $SL(2,\bbR)$-invariant function of the scalars, to be determined later, $L_0$ is the
Born-Infeld function \eq{3.22} and $\ve_{(p+1)}$ is the bosonic volume form,

\be
 \ve_{(p+1)}:=\frac{1}{(p+1)!} E^{a_{p+1}}\ldots E^{a_1} \ve_{a_1\ldots a_{p+1}}\ .
 \la{5.2}
\ee

We shall only need to show that

\be
 W_{p,2}=(d L_{DBI})_{p,2}\ .
 \la{5.3}
\ee

A short calculation yields

\be
 (d L_{DBI})_{p,2}=-\frac{i}{2} fL_0 \ve_a E^\b E^\a ( (h\c^a + \c^a (h^{-1})^T) h^T)_{\a\b} \ ,
 \la{5.5}
\ee

where

\be
 \ve_a:=\frac{1}{p!} E^{b_p}\ldots E^{b_1} \ve_{a b_1\ldots b_p}\ .
 \la{5.5.1}
\ee

Pulling back the dimension zero $G$s and concentrating on the terms with $E^\b E^\a$ we find

\be
 (\sum G^{(2n+1)})_{p,2} = -i E^\b E^\a\sum ((a')^{n-2} \c^{(2n-1)}h^T)_{\a\b}\ .
 \la{5.6}
\ee

Thus we have

$$
\left(e^{-\cF} \sum G\right)_{p,2} =\qquad\qquad\qquad\qquad\qquad\qquad\qquad
\qquad\qquad\qquad\qquad\qquad\qquad\qquad\qquad\qquad\nonumber
$$
\bea
 &=&-i E^\b E^\a \sum \left(\frac{(-1)^m (a')^{n-2}}{2^m m!(2n-1)!}E^{a_{2m}}\ldots
 E^{a_1} E^{b_{2n-1}}\ldots E^{b_1}(\c_{b_1\ldots b_{2n-1}}h^T)_{\a\b}\cF_{a_1\ldots a_{2m}} \right)\nn\w2
 &=& i E^\b E^\a \sum \frac{(-1)^m(a')^{n-2}}{2^m m!(2n-1)! }\ve_c \ve^{ca_1\ldots a_{2m}
 b_1\ldots b_{2n-1}}
 (\c_{b_1\ldots b_{2n-1}}h^T)_{\a\b} \cF_{a_1\ldots a_{2m}} \nn\w2
 &=& iE^\b E^\a\sum \frac{(a')^{n-2}}{2^m m!}\ve_a (\c^{a b_1\ldots b_{2m}}\c_{(p+1)}h^T)_{\a\b}
 \cF_{b_1\ldots b_{2m}}\ ,
 \la{5.7}
\eea

where $\cF_{a_1\ldots a_{2m}}:= \cF_{[a_1 a_2}\ldots \cF_{a_{2m-1}a_{2m}]}$, and where $2m+2n=p+1$.
Writing $\c^{2m+1}=\half\{\c,\c^{2m}\}$ and using the explicit formula for $h$ we find

\be
 \left(e^{-\cF} \sum G\right)_{p,2}= -\frac{i}{2} f L_0
 \ve_a E^\b E^\a ( (h\c^a + \c^a (h^{-1})^T) h^T)_{\a\b}\ ,
 \la{5.8}
\ee

which is what we wanted to show. The function $f$ is determined to be

\be
 f=(a')^{\frac{p-3}{2}}\ .
 \la{5.8.1}
\ee

This is in agreement with \cite{Bergshoeff:2006gs}, although in the current approach it is determined by supersymmetry. On the face of it, there appears to be a conflict in the string case with the tension formula given there, but this turns out to be due to the fact that the universal formula given here goes with the action that contains a Born-Infeld field even for $p=1$. We shall see in section 5 that when this is eliminated this apparent disagreement disappears.

In conclusion we have shown that the Green-Schwarz action

\be
 S = \int_{M_0}\, (L_{DBI} - L_{WZ})\ ,
 \la{5.9}
\ee

where

\be
 L_{DBI}=(a')^{\frac{p-3}{2}}\sqrt{-\det\, (\h_{ab} + \cF'_{ab})}\,\ve_{(p+1)}\ ,
 \la{5.9.1}
\ee

and

\be
 L_{WZ}=(e^{-\cF} \sum C^{(2n)})_{p+1}\ ,
 \la{5.10}
\ee

and where the integrand is to be interpreted as in \eq{3.23}, is covariant under kappa-symmetry. It
is also manifestly $SL(2,\bbR)$-invariant.


\section{$(p,q)$ Strings}


If we specialise to the case of strings, the above construction gives the $SL(2,\bbR)$ covariant string action,

\begin{equation}
S=\int d^2 x\,\sqrt{-\det g_{\mathrm
E}}\left((a')^{-1}\sqrt{1-\left(\frac{F}{a'}\right)^2}+\frac{b}{a}F\right)-\int C^{(2)}\,,
\end{equation}

where we have expanded the determinant and defined $F:=\mathcal{F}_{01}$. The equation of motion for
the gauge field gives

\begin{equation}
(a')^{-3}\left(1-\left(\frac{F}{a'}\right)^2\right)^{-1/2}F+\frac{b}{a}=-k\,,
\end{equation}

where $k$ is a constant which can be absorbed by shifting $b\rightarrow b+ka$. This in turn can be
seen to be equivalent to shifing $\tilde q^R\rightarrow\tilde q^R+kq^R$ (see below). We will
therefore set $k=0$ and the above equation becomes

\begin{equation}
F=-\frac{ab}{2}\sqrt{(a')^2-F^2}\,.
\end{equation}

Solving for $F$ we get

\begin{equation}
F=\frac{\sqrt{2}(a')^2b}{\sqrt{a^2b^2+4}}\ .
\end{equation}

For the terms involving $F$ in the action we thus get

\begin{equation}
(a')^{-1}\sqrt{1-\left(\frac{F}{a'}\right)^2}+\frac{b}{a}F =(a^2b^2+4)\frac{F}{a^3b}
=\frac{1}{2a'}\sqrt{a^2b^2+4}\,.
\end{equation}

Using the fact that

\begin{equation}
b^2=\tilde q\mathcal M\tilde q^{\mathrm{T}}-4/a^2
\end{equation}

we get

\begin{equation}
(a')^{-1}\sqrt{1-\left(\frac{F}{a'}\right)^2}+\frac{b}{a}F =-\frac{1}{\sqrt2}\sqrt{\tilde q\mathcal
M\tilde q^{\mathrm{T}}}\,.
\end{equation}

In the physical gauge $\mathcal M$ is given by

\begin{equation}
\mathcal M=\frac{1}{\tau_2} \left(\begin{array}{cc}
1&-\tau_1\\
-\tau_1&|\tau|^2
\end{array}\right)
=e^\phi\left(\begin{array}{cc}
1&-C_0\\
-C_0&C_0^2+e^{-2\phi}
\end{array}\right)\ ,
\la{M}
\end{equation}

where $\t:=C_0+ie^{-\f}$. Taking $\tilde q^R=\sqrt2(p,q)$ (this gives $q^R=-\frac{\sqrt2}{(p^2+q^2)}(-q,p)$), where $p,q$ here denote non-negative integers, we find

\begin{equation}
\tilde q\mathcal M\tilde q^{\mathrm{T}}=2e^\phi(q^2-2pqC_0+p^2C_0+p^2e^{-2\phi}))
=2(e^\phi(p-qC_0)^2+q^2e^{-\phi})\,.
\end{equation}

Thus the action becomes

\begin{equation}
S=-\int d^2 x\,\sqrt{(e^\phi(p-qC_0)^2+q^2e^{-\phi})}\sqrt{-\det g_{\mathrm E}}-\int
C^{(2)}\,,
\end{equation}

which contains the appropriate expression for the tension of a $(p,q)$-string in the Einstein
frame \cite{Schwarz:1995dk}. The Wess-Zumino term satisfies

\begin{equation}
dC^{(2)}=G^{(3)}-H\frac{b}{a}=-\tilde q^R F_R^{(3)}=-\sqrt{2}(pF_1^{(3)}+qF_2^{(3)})\,,
\end{equation}

which reduces to the correct couplings for an F- or D-string when $(p,q)$ is $(1,0)$ or $(0,1)$.


\section{Concluding remarks}


In this article we have completed IIB supergravity theory in superspace by incorporating all of the forms including the field strengths corresponding to the ten-form potentials introduced in \cite{Bergshoeff:2005ac}. We then went on to use the superembedding formalism to construct Green-Schwarz actions which are invariant under $SL(2,\bbR)$ as well as kappa symmetry. Given the basic superembedding constraints, \eq{3.6} and \eq{3.7}, this formalism, combined with the superform method of constructing component actions from superspace, then determines the desired actions in a systematic fashion.

The D-brane actions given here can be thought of as those that lie within the $SL(2,\bbR)$ orbit of the usual D-brane actions. An interesting question is whether there might be others. For example, it is known that there are additional D7-brane solutions of IIB supergravity which are half-supersymmetric \cite{Bergshoeff:2006jj}. These branes correspond to couplings to a nine-form field strength of the form $q^{RS} F^{(9)}_{RS}$, where the matrix $q^{RS}$ is non-singular. Another interesting question is whether the doublet of eleven-form field strengths have any significance for brane physics.

It would be interesting to extend the results of this paper to other maximal supergravity theories  where the complete sets of potentials are also known \cite{Riccioni:2007au,Bergshoeff:2007qi,Bergshoeff:2006qw}. It may also be possible to extend the formalism to the non-abelian case in the boundary fermion formalism \cite{Howe:2005jz,Howe:2007eb}, at least at the level of classical fermions.

\pagebreak


\section*{Acknowledgements}


We thank James Drummond, Mees de Roo and Dima Sorokin for interesting discussions. S.K. would like to thank the Centre for Theoretical Physics at the University of Groningen for hospitality.

E.B.~is supported by the European Commission FP6 program
MRTN-CT-2004-005104 in which E.B.~is associated to Utrecht University. The work of E.B.~is partially supported by the Spanish grant BFM2003-01090 and  by a Breedte Strategie grant of the University of Groningen. This work was also supported in part by EU-grant (Superstring Theory) MRTN-2004-512194.

\appendix

\section{Conventions}


The $SO(2)$ formalism in this paper is related to the $U(1)$ formalism of \cite{Howe:1983sr} in the following way (see also \cite{Berkovits:2001ue}). For a $Spin(2)$ vector $y$ one writes

\be
 y^{\pm}=\frac{1}{\sqrt{2}}(y^1\pm iy^2)\ ,
 \la{A.1}
\ee

while for covectors one has

\be
 y_{\pm}=\frac{1}{\sqrt{2}}(y^1\mp iy^2)\ .
 \la{A.2}
\ee

Thus the metric is off-diagonal in the complex basis,

\be
 \d^{+-}=\d_{+-}=1\ .
 \la{A.3}
\ee

The same conventions are used for $SO(2)$ vectors except that each index now carries double the charge; for example, a vector $Y^r$ has components $(Y^{++},Y^{--})$ in the complex basis. The components of the $\ve$-tensors, $\ve_{ij}$ and $\ve_{rs}$ are specified by

\be
 \ve_{+-}=\ve_{++--}=i\ .
 \la{A.4}
\ee

In order to compare with  \cite{Howe:1983sr} it is necessary to change $E^a$ to $-E^a$ and to change the signs of $F^{(3)}$ and $F^{(5)}$. In this paper the summation convention for odd indices is

\be
 \chi^{\a i} \r_{\a i}=\chi^{\a +}\r_{\a +} +\chi^{\a -}\r_{\a -}\ ,
 \la{A.5}
\ee

whereas there is a minus sign for the second term in  \cite{Howe:1983sr}.

We have taken the matrix of scalars, in the physical gauge and in a real basis, to be

\be
 \cU=\frac{1}{\sqrt{\t_2}}\left(\ba{cc} \t_2& 0\\ \t_1 & 1\ea\right)\ ,
 \la{A.6}
\ee

where $\t:=C_0+ie^{-\f}$. This then gives $\cM$ as in \eq{M}.


\section{Proof that $H_t^{p,2}=0$.}


In the text it is stated that the cohomology group $H_t^{p,2}=0$ for any D$p$-brane. In this appendix we sketch the proof. We wish to show that the equation $t_0 W_{p,2}=0$ only has cohomologically trivial solutions of the form $W_{p,2}=t_0 K_{p+1,0}$. Let $T$ denote the dimension zero torsion on the brane considered as a vector-valued $(0,2)$-form and let $W$ be the vector-valued $(0,2)$-form obtained from $W_{p,2}$ by dualising on the even indices. The equation to be solved can then be written in the form

\be
 T^{[a}_{(\a\b} W^{b]}_{\c\d)}=0\ ,
 \la{B.1}
\ee

which we shall abbreviate to $T \xz W=0$, and we want to show that $W=TK$ for some function $K$ which is the dual of $K_{p+1,0}$. Note that $T\xz W=-W\xz T$. We first show that the cohomology of $T=\c$ is trivial; to do this one expands $W$ in a basis of symmetric ten-dimensional gamma-matrices and then goes through \eq{B.1} systematically one representation at a time. It is straightforward to show that $W=\c K$ for some function $K$.

The second step is to extend this result to the full theory on the brane, for which $T$ is given by \eq{3.14}, perturbatively; that is, we expand both $T$ and $W$ in powers of $\cF_{ab}$. Note that this is the only possibility since $\cF_{ab}$ is the only covariant dimension zero field in the problem. Thus we write

\be
 T=T_0 +  T_1 +\ldots\ ,
 \la{B.2}
\ee

and similarly for $W$ and $K$, although the latter has only even terms. We can normalise $T$ such that $T_0=\c$ so that we have $W_0=T_0 K_0$, where we could choose $K_0=1$ if desired. The proof is by induction; let us suppose that it holds up to the $n$th order, i.e.

\be
 W_n=\sum_{k=0}^{n} T_k K_{n-k}\ ,
 \la{B.3}
\ee

then at the $(n+1)$th order we have

\be
 T_0\xz W_{n+1} + \sum_{k=1}^{n+1} T_k\xz W_{n+1-k}=0\ .
 \la{B.4}
\ee

Using \eq{B.3} we can write the sum as

\bea
 \sum_{k=1}^{n+1} T_k\xz W_{n+1-k}&=&\sum_{k=1}^{n+1} T_k \xz \sum_{l=0}^{n+1-k} T_l K_{n+1-(k+l)}\nn\w1
 &=&\left(\sum_{k=1}^{n+1} T_k \xz \sum_{l=1}^{n+1-k} T_l K_{n+1-(k+l)}\right)-
 \left(T_0\xz \sum_{k=1}^{n+1} T_k K_{n+1-k}\right)\ .
\la{B.5}
\eea

In the first term on the right in the last line the sum over $l$ can be extended up to $n+1$ if we define $K_m=0$ for $m$ negative. We can then easily see that it vanishes by symmetry since $T_k\xz T_l=-T_l\xz T_k$. Equation \eq{B.4} therefore becomes

\be
 T_0\xz \left(W_{n+1}-\sum_{k=1}^{n+1} T_k K_{n+1-k}\right)=0\ ,
 \la{B.6}
\ee

which has solution

\be
 W_{n+1}=\sum_{k=0}^{n+1} T_k K_{n+1-k}\ ,
 \la{B.7}
\ee

as we have already argued that the cohomology of $T_0$ is trivial. And this completes the proof.


\section{The $SL(2,\bbZ)$-covariant superstring}


The $SL(2,\bbZ)$-covariant superstring action of Cederwall-Townsend \cite{Cederwall:1997ts} is, in $SO(2)$ notation,

\begin{equation}
S=\frac{1}{2}\int d^2 x\,\lambda\left(\det g_{\mathrm E}+2\mathcal F_r \delta^{rs}\mathcal F_s\right)\,,
\end{equation}

where $g_{\mathrm E}$ is the metric in the Einstein frame, $\l$ is an auxiliary scalar field and $\mathcal F_r=\frac{1}{2}\varepsilon^{mn}\mathcal F_{rmn}$, $m,n=0,1$ being worldsheet coordinate indices. The doublet of modified field strengths is defined as

\begin{equation}
\mathcal F_r=\mathcal U_r{}^R(d\cA_R-B_R)\,,
\end{equation}

where $B_R$ is the potential for $F^{(3)}_R$. We wish to integrate out one of the gauge-fields $A_R$ and show that this reproduces the action given in  \eq{5.9}, for the case $p=1$.

We start by expanding $\mathcal F_r$ in the orthonormal basis $(V^r,\tilde V^r)$;

\begin{equation}
\mathcal F_r=V_r(V^s\mathcal F_s)+\tilde V_r(\tilde V^s\mathcal F_s)\,.
\end{equation}

Then we use \eq{3.2} to express $V_r$ ($\tilde V_r$) in terms of $q^r$ ($\tilde q^r$). The expression in the Lagrangian becomes

\begin{equation}
2\mathcal F_r \delta^{rs}\mathcal F_s=2((V^r\mathcal F_r )^2+(\tilde V^r\mathcal F_r)^2)
=\frac{2}{a^2}\mathcal F^2+\frac{1}{2}(-a\tilde{\mathcal F}+b\mathcal F)^2\,,
\end{equation}

where $q^r\mathcal F_r=\mathcal F$ and similarly for $\tilde{\mathcal F}$.

In order to integrate out the gauge-field corresponding to $\tilde{\mathcal F}$ we define $\tilde F=\tilde q^R\varepsilon^{mn}\partial_m \cA_{Rn}=\varepsilon^{mn}\partial_m\tilde \cA_n$, so that $\tilde{\mathcal F}=\tilde F-\tilde q^rB_r$. To treat $\tilde F$ as independent we add the term

\begin{equation}
\mu(\tilde F-\varepsilon^{mn}\partial_m\tilde \cA_n)=\mu(\tilde{\mathcal F}+\tilde q^rB_r-\varepsilon^{mn}\partial_m\tilde \cA_n)
\end{equation}

to the Lagrangian, where $\mu$ is a Lagrange multiplier enforcing the constraint that $\tilde F$ be closed. We can now set the variation of the action with respect to $\tilde F$ to zero. This gives the equation

\begin{equation}
-a\tilde{\mathcal F}+b\mathcal F=\frac{2\mu}{a\lambda}\,.
\end{equation}

The equation of motion for $\tilde \cA$ simply implies that $\mu$ is constant. Plugging these results into the action and collecting terms according to their dependence on $\lambda$ we get

\begin{equation}
S=\int d^2 x\,\left(\frac{\lambda}{2}\left(\det g_{\mathrm E}+\mathcal F'{}^2\right)
-\frac{\mu^2}{a^2\lambda}+\mu\left(\frac{b}{a}\mathcal F+\tilde q^r B_r\right)
\right)\,.
\end{equation}

Finally we eliminate $\lambda$ by its equation of motion

\begin{equation}
\lambda=\frac{\mu}{a'}\left(-\det g_{\mathrm E}-\mathcal F'{}^2\right)^{-1/2}
\end{equation}

and we get

\begin{equation}
S=\int d^2 x \,\left(-\frac{\mu}{a'}\sqrt{-\det g_{\mathrm E}-\mathcal F'{}^2}
+\mu\left(\frac{b}{a}\mathcal F+\tilde q^r B_r\right)
\right)\,.
\end{equation}

Taking the constant $\mu$ to be $-1$, noting from \eq{4.3} that $\tilde q^r B_r=-\frac{1}{2}\varepsilon^{mn}C^{(2)}_{mn}$ and writing

\begin{equation}
\mathcal F'=\frac{1}{2}\varepsilon^{mn}\mathcal F'_{mn}=\sqrt{-g_{\mathrm E}}\,\frac{1}{2}\varepsilon^{ab}\mathcal F'_{ab}\,,
\end{equation}

we see that the action can be written as

\begin{equation}
S=\int d^2 x \,\sqrt{-\det g_{\mathrm E}}\,(a')^{-1}\sqrt{1-\left(\frac{1}{2}\varepsilon^{ab}\mathcal F'_{ab}\right)^2}
+\int\,(-C^{(0)}\mathcal F+C^{(2)})\,.
\end{equation}

This is easily seen to be precisely the action given in \eq{5.9} when $p=1$ by expansion of the determinant in the DBI--term. This completes the proof of the equivalence of the two actions.


\end{document}